\documentclass[aps,prl,twocolumn,amsmath,amssymb, superscriptaddress,tightenlines]{revtex4}
\usepackage{graphicx}
\usepackage{epsfig}
\usepackage{dcolumn}
\usepackage{bm}
\usepackage{amssymb}
\usepackage{bm}
\usepackage{amsmath, bm}
\usepackage{upgreek}
\usepackage[colorlinks,citecolor=blue,linkcolor=blue,hyperindex]{hyperref}

\begin{document}

\title{Thermal equilibration on the edges of topological liquids}
\author{Ken K. W. Ma and D. E. Feldman}
\affiliation{Brown Theoretical Physics Center and Department of Physics, Brown University, Providence, Rhode Island 02912, USA}
\date{\today}


\begin{abstract}
Thermal conductance has emerged as a powerful probe of topological order in the quantum Hall effect and beyond. The interpretation of experiments crucially depends on the ratio of the sample size and the equilibration length, on which energy exchange among contra-propagating chiral modes becomes significant. We show that at low temperatures the equilibration length
diverges as $1/T^2$  for almost all Abelian and non-Abelian topological orders. A faster $1/T^4$ divergence is present on the edges of the non-Abelian PH-Pfaffian and negative-flux Read-Rezayi liquids. We address experimental consequences of the $1/T^2$ and $1/T^4$ laws in a sample, shorter than the equilibration length. 
\end{abstract}

\maketitle


\textbf{Introduction} - The universal properties of a gapped state of matter are known as its topological order~\cite{Wen_book}. In the language of bulk physics, the order reduces to the list of possible anyons.  In terms of the edge, the order tells about gapless chiral edge modes. Topologically protected edge modes give rise to quantized transport. In particular, a quantized electrical conductance is a defining feature of the fractional quantum Hall (FQH) effect. Yet, the electrical conductance alone is insufficient to determine the order. For example, all orders of Kitaev's sixteen-fold way~\cite{kitaev2006,16-fold} exhibit the same conductance. Which of them is present at the filling factor  $5/2$  in GaAs is hotly debated~\cite{16-fold,MFreview}. 
Moreover, electric transport is not a useful probe of magnetic materials, such as candidate Kitaev spin liquids~\cite{kitaev2006}.

Another quantized transport coefficient is the thermal conductance~\cite{Kane_thermal,Read-Green,Cappeli}. It gives way more information than the electrical conductance but is also harder to measure~\cite{MFreview}. Only very recently have thermal-conductance data become available in FQHE~\cite{Banerjee2017,Banerjee2018} and RuCl$_3$~\cite{RuCl3}. 
The interpretation of such data is straightforward on chiral edges, where all modes propagate in the same direction and have the same temperature.
If both up- and down-stream contra-propagating modes are present, the quantization depends on equilibration among the modes~\cite{Banerjee2017,so-2018,Ken2019}. This does not pose a challenge for the electric transport since the observed voltage equilibration length is believed to be on the order of a micron or shorter~\cite{Lafont2019}. The temperature equilibration length $\ell_{\rm eq}$ extends to tens of microns and can be comparable to or greater than the size of a mesoscopic device~\cite{Banerjee2017}. As a result, large finite-size effects have been observed in low-temperature thermal transport. Hence, the interpretation of the data requires the knowledge of the temperature dependence of the equilibration length. We find it in this paper in the limit of low temperatures.

The results are remarkably universal. We predict the same $1/T^2$ dependence for almost all Abelian and non-Abelian orders. Exceptions from the $1/T^2$-law are the PH-Pfaffian order~\cite{Son2015,Zucker2016,APf_Lee2007,FCV,BNQ} and a family of orders~\cite{Jolicoeur2007}, 
related to the Read-Rezayi (RR) states~\cite{RR}, where a different universal $1/T^4$ dependence is found. The universality comes from strong restrictions on possible scaling dimensions of Bose operators responsible for energy exchange between the up- and down-stream modes in the edge conformal field theory (CFT). We address an experimental test of the predicted universality in a geometry with an edge, shorter than $\ell_{\rm eq}$. At a small number of filling factors such as $\nu=\frac{3n-1}{4n-1}$, $n>1$, the predicted low-temperature behavior of the equilibration length is non-universal.

\begin{figure} [htb]
\includegraphics[width=3.0in]{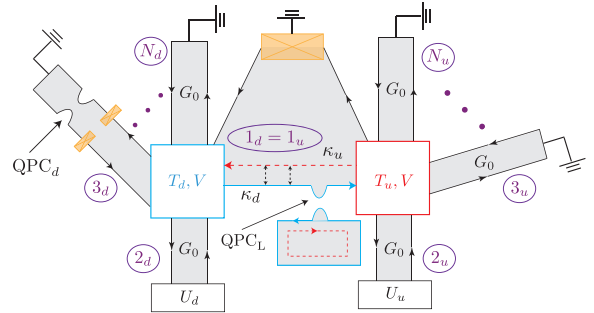}
\caption{Thermal equilibration of downstream (solid blue line) and upstream (dashed red) channels is investigated. Gray areas contain FQH liquid. White and yellow boxes are floating contacts.}
\label{fig:creation}
\end{figure}

\textbf{Thermal equilibration length} - Three length scales are important: the equilibration length $\ell_{\rm eq}$, the edge length $L$, and the thermal length $\ell_{\rm th}\sim \hbar v/T$, where $v$ is on the order of the velocities of the edge modes. The physics at the scales $<\ell_{\rm th}$
 is dominated by quantum interference and does not depend on the temperature. The upstream and downstream modes emerge from two reservoirs (Fig. 1) at the temperatures $T_{\rm u}$ and $T_{\rm d}$ and exchange heat. We assume
that $\ell_{\rm th}\ll L \ll \ell_{\rm eq}$. 
This assumption means that $T_{\rm u,d}$ remain approximately constant along the edge and allows treating thermal exchange perturbatively.
For the convenience of the scaling analysis we will assume that the ratio $T_{\rm u}/T_{\rm d}$ remains fixed in the limit $T\rightarrow 0$.

Down- and up-stream modes are coupled by random and non-random interactions. When contra-propagating edge modes have a temperature difference $\Delta T$, a heat flux of $\kappa_0 T\Delta T$ flows between them on a length segment $\ell_{\rm eq}$, where $\kappa_0=\pi^2k^2_{\rm B}/3h$ defines a thermal conductance quantum $\kappa_0T$. We employ the renormalization group (RG) technique to analyze the temperature dependence of $\ell_{\rm eq}$.

\textit{Random interaction} 
is included in the Hamiltonian as
\begin{eqnarray}
\label{dima-2}
H_{\rm r}=\int_0^L \xi(x)\hat{O}(x)~ dx.
\end{eqnarray}
Here, $\xi(x)$ is random, and $\hat{O}(x)$ is a product of two operators acting on upstream and downstream modes. The explicit form of $\hat{O}(x)$ depends on the edge structure of the system, but the operator must always be bosonic. We assume that the correlation length of the random potential is shorter than 
$\ell_{\rm th}$ and set $\langle \xi(x)\xi(x')\rangle =W\delta(x-x')$. 

We will see that $H_{\rm r}$ is typically irrelevant in the RG sense. Hence, it is sufficient to use
the leading order RG equation for $W$ ~\cite{Kane-disorder-dominated}
\begin{eqnarray} \label{eq:RG-eqn}
\frac{dW}{d\ell}
=(3-2\Delta)W,
\end{eqnarray}
where $\Delta$ is the scaling dimension of the operator $\hat{O}$. The RG flow stops at 
$\ell_{\rm th}$. From Eq.~\eqref{eq:RG-eqn}, the effective coupling (after renormalization) takes the form $W_{\rm eff}\sim W \ell_{\rm th}^{3-2\Delta}$. 
The energy flux per unit length between the up- and down-stream modes can be found from the Fermi's golden rule:
\begin{eqnarray}
\label{dima-old-3}
\mathcal{J}
\sim \frac{W_{\rm eff}}{\ell_{\rm th}}\left(\kappa_0 T\Delta T\right)
\sim W\ell_{\rm th}^{2-2\Delta}\left(\kappa_0 T\Delta T\right).
\end{eqnarray}
The thermal equilibration length $\ell_{\rm eq}$ is defined by 
$\kappa_0 T\Delta T\sim \mathcal{J}\ell_{\rm eq}$ so that
\begin{eqnarray} \label{eq:l-eff-RG}
\ell_{\rm eq}
\sim T^{2-2\Delta}.
\end{eqnarray}
Thus, our problem reduces to computing $\Delta$, that is, finding the most relevant possible operator $\hat O(x)$, Eq. (\ref{dima-2}).

{\it Non-random interactions} 
$H_{\rm N}=\xi\int_0^L \hat{O}(x)~ dx$
can be analyzed in a similar way \cite{dima-footnote-03-19}.  However, only operators $\hat O(x)$ that couple at least three one-dimensional edge modes with three different velocities should be considered. Otherwise, it is impossible to conserve both momentum and energy in a scattering event. Besides, non-random operators that transfer 
a finite charge between 
modes are ineffective in equilibration due to the momentum mismatch between the modes \cite{Dima_comment2018}. Indeed, charge tunneling in a magnetic field involves a finite momentum boost, which is incompatible with the energy conservation for low-energy excitations at $T\rightarrow 0$. Thus, the list of non-random operators that can transfer energy between the modes
is very limited, and in all cases, the equilibration length turns out to be determined by random operators (\ref{dima-2}).

\textbf{Jain states} -  A generic Lagrangian density of an edge of an Abelian liquid at the filling factor $\nu$ without disorder \cite{Wen_book} is
\begin{align}
\label{dima-new-1}
\nonumber
L=&~\frac{1}{4\pi\nu}[\partial_t\phi_c\partial_x\phi_c-v_c(\partial_x\phi_c)^2]-\sum_i w_i\partial_x\phi_c\partial_x\phi_i
\\
&+\sum_{ij}\frac{1}{4\pi}[{\tilde K_{ij}}\partial_t\phi_i\partial_x\phi_j-V_{ij}\partial_x\phi_i\partial_x\phi_j],
\end{align}
where the charge mode $\phi_c$ describes the total charge density $e\partial_x\phi_c/2\pi$ on the edge and $\phi_i$ are the neutral modes. $v_c$ is the speed of the downstream charge mode. $\tilde K$ determines the commutators of the neutral mode operators and $V$ determines the velocities of the neutral modes. Their propagation directions depend on the signs of the eigenvalues of 
$\tilde K$ and are upstream for negative eigenvalues.
We focus on edges defined by chemical etching and assume that any screening gates are far away. Since the charge mode participates in the long-range Coulomb interaction and the neutral modes do not, we expect $v_c$ to be much greater than the velocities of the neutral modes and the intermode interactions $V_{ij}$ and $w_i$. 

The Jain states \cite{Jain_book} possess upstream edge modes at the filling factors $\nu=\frac{n}{2np-1}$ with $n>1$. As was discussed by Kane and Fisher \cite{Kane-disorder-dominated}, disorder gives rise to charge tunneling operators in the action. In our limit of large $v_c$, those operators are always relevant in the RG sense. They are responsible for voltage equilibration. Following Kane and Fisher, we change variables in the action so that those operators disappear. In the new variables, the $(n-1)$ upstream neutral modes propagate at the same speed and exhibit the SU$(n)$ symmetry. Their interaction with the charge mode involves only irrelevant random couplings (\ref{dima-2}) with $\hat O=\hat O_c\hat O_n$ \cite{foot2}, where $\hat O_{c,n}$ act on the charge and neutral modes. The conservation of the electric charge 
demands that $\hat O_c$ be a combination of derivatives of $\phi_c$. Any such operator is bosonic. Since $\hat O$ must satisfy the Bose statistics, $\hat O_n$ is also bosonic. We next observe that upstream and downstream modes are described by two chiral CFTs. The scaling dimensions of bosonic operators in chiral CFTs are integers, which must be greater than zero 
unless the operators are trivial constants. It follows that the scaling dimensions of $\hat O_{n,c}$ cannot be less than 1 and hence $\Delta\ge 2$. Operators with  $\Delta=2$ do exist, e.g., $\hat O=\partial_x\phi_c\partial_x\phi_i$.  Thus, $\ell_{\rm eq}\sim 1/T^2$ from Eq. (\ref{eq:l-eff-RG}).

\textbf{Non-Abelian orders} - We first consider topological orders of the sixteen-fold way at half-integer filling factors \cite{16-fold}. These orders include the non-Abelian Moore-Read Pfaffian state and  its two non-Abelian relatives, anti-Pfaffian and PH-Pfaffian, which are seen as the leading candidates at $\nu=5/2$. 
Five more orders of the sixteen-fold way are non-Abelian, and eight orders are Abelian. In all cases, the edge theory has the same general structure \cite{16-fold} with one downstream charged boson and several neutral Majorana modes. It is conventional in the literature to combine pairs of Majoranas into neutral bosons, but we will not do so. 
The number of the Majorana modes is even for Abelian orders and odd for non-Abelian orders.
In the absence of disorder, the edge Lagrangian density contains a bosonic charge field $\phi_c$ and Majorana fermions $\psi_k$:

\begin{align}
\label{dima-new-2}
\nonumber
L=&~\frac{2}{4\pi}\partial_x\phi_c(\partial_t-v_c\partial_x)\phi_c 
\\
&+\sum_{k=1}^{|C|}i\psi_k(\partial_t-u~{\rm sign} [C]\partial_x)\psi_k,
\end{align}
where the Chern number $C$ sets the number $|C|$ of the neutral Majoranas, $k=1,\dots,|C|$. 

The Pfaffian order corresponds to $C=1$. This reflects a single Majorana co-propagating downstream with the charged mode. We are not interested in the Pfaffian state and any other order with $C\ge 0$ since all modes have the same temperature automatically at $C\ge 0$. $C=-3$ in the anti-Pfaffian state, which is of interest for this paper. 
{\it A priori}, the velocities of the $|C|$ Majoranas do not have to be the same. As discussed in Refs. \cite{APf_Lee2007,APf_Levin2007}, random charge tunneling, responsible for charge equilibration, makes the velocities of the Majorana modes identical. This comes from exactly the  same mechanism as in the Jain states \cite{Kane-disorder-dominated}.
A similar physical picture \cite{guang2013} applies for other $C<-1$: all upstream Majoranas propagate at the same speed. Some subtleties \cite{guang2013,guang2014} are present at $C=-4$ (the anti-331 state) and $C=-2$ (the 113 state). In particular, the anti-331 edge action contains a four-fermion contribution $\sim \psi_1\psi_2\psi_3\psi_4$. This, however, does not affect the discussion below. On the other hand, the physics of the PH-Pfaffian order with $C=-1$ ends up entirely different since only one neutral Majorana is present on the edge. We will start with the case of $C<-1$.


The argument from the discussion of the Jain states applies almost verbatim at $C<-1$. The only change is a different choice of an operator $\hat O$ with $\Delta=2$: $\hat O=i\partial_x\phi_c\psi_n\psi_m$, $n\ne m$, where the imaginary-unit factor ensures the hermiticity of $\hat O$. Thus, $\ell_{\rm eq}\sim 1/T^2$ again. 

We now turn to the anti-RR states at $\nu=2/(k+2),~k>2$. Their edge structures are addressed in detail in Ref. \cite{Bishara_PH_RR}. The bosonic charge mode $\phi_c$ runs downstream; the upstream neutral modes have identical velocities and are described by the SU$(2)_k$ chiral Wess-Zumino-Witten model \cite{cft}. All states in that model can be organized into representations of the affine SU$(2)$ algebra of the currents $J^x, ~J^y,~ J^z$ at level $k$ \cite{cft}. The expressions for the currents can be found in Ref. \cite{Bishara_PH_RR}. They are Bose operators of scaling dimension 1. To compute $\ell_{\rm eq}$ we repeat again the same argument as for the Jain states and just change the expression for $\hat O=\partial_x\phi_c J^\alpha$. Again, $\ell_{\rm eq}\sim 1/T^2$. 

At this point, one might think that the $1/T^2$ scaling is a general rule. Yet, exceptions exist. The simplest exception is the PH-Pfaffian order \cite{Son2015,Zucker2016,APf_Lee2007,FCV,BNQ} with the edge structure (\ref{dima-new-2}) with $C=-1$. 
There is only one upstream neutral Majorana mode $\psi$. One easily sees that no local Bose operator  $\hat O=\partial_x \phi_c\hat O_n$ with scaling dimension 2 can be found. 
Indeed, we do not have enough fields to build a product of two Majoranas, as in the other states of the sixteen-fold way. The square of a single Majorana is a trivial constant. This leaves us with  $\hat O_n=i\psi\partial_x\psi$ as the option, most relevant \cite{foot3} in the RG sense. 
It follows that $\Delta=3$ and $\ell_{\rm eq}\sim 1/T^4$. Unusually rapid growth of the equilibration length at low temperatures in the PH-Pfaffian state is of interest \cite{Banerjee2018} for the interpretation of recent thermal conductance data.

The PH-Pfaffian order is a relative of the negative-flux RR states \cite{Jolicoeur2007} at the filling factors $\nu=\frac{k}{3k-2}$, $k>2$. The edge contains a downstream Bose charge mode and an upstream sector described by the SU$(2)_k/$U$(1)$ coset chiral CFT \cite{dima-SB} also known as the parafermion theory. 
As shown in Supplemental Material \cite{SM-dima}, the equilibration length follows  the $\ell_{\rm eq}\sim 1/T^4$ law for those states. To demonstrate that, we prove that the upstream sector has no spin-1 fields mutually local with all primaries, just like no such field exists in the PH-Pfaffian state.

\textbf{Anti-Jain states} - The particle-hole conjugates of the Jain states occur at $\nu=1-\frac{n}{2np\pm1}$.
We start with $\nu=1-\frac{n}{2np+1}$ and only consider $p>1$ since $\nu=1-\frac{n}{2n+1}=\frac{n+1}{2(n+1)-1}$ were addressed above. The Lagrangian density of the edge without disorder is

\begin{align}
\label{dima-new-5}
\nonumber
L=&~\frac{\partial_x\phi_1(\partial_t-v_1\partial_x)\phi_1}{4\pi}
-\frac{\partial_x\phi_\rho(\partial_t+v_\rho\partial_x)\phi_\rho}{4\pi(1-\nu)}
\\
&~-w\partial_x\phi_1\partial_x\phi_\rho+L_n,
\end{align} 
where $\phi_1$ is a downstream charge field with the charge density $e\partial_x\phi_1/2\pi$, $\phi_\rho$ is an upstream charge field with the charge density $e\partial_x\phi_\rho/2\pi$, and $L_n$ describes 
$(n-1)$ upstream neutral modes, moving at the same velocity by the Kane-Fisher mechanism \cite{Kane-disorder-dominated}.  The interaction $w$ is strong, and we switch to the overall downstream charge mode
 $\phi_c=\phi_\rho+\phi_1$ and an additional upstream neutral mode $\phi_n=\phi_1+\frac{\phi_\rho}{1-\nu}$. The only downstream mode $\phi_c$ is much faster than all other modes. The Lagrangian density becomes

\begin{align}
\label{dima-new-6}
\nonumber
L=&~\frac{\partial_x\phi_c(\partial_t-v_c\partial_x)\phi_c}{4\pi\nu}
-\frac{1-\nu}{4\pi\nu}\partial_x\phi_n(\partial_t+v_n\partial_x)\phi_n
\\
&-\tilde w\partial_x\phi_n\partial_x\phi_c+L_n,
\end{align} 
where $\tilde w, v_n \ll v_c$.  The coupling $\tilde w$ disappears in the language of the normal modes $\tilde\phi_{n,c}$:

\begin{align}
\label{dima-new-7}
\phi_c=&~\tilde\phi_c\cosh\theta+\tilde\phi_n\sqrt{1-\nu}\sinh\theta;\\
\label{dima-new-8}
\phi_n=&~\tilde\phi_n\cosh\theta+\tilde\phi_c\frac{\sinh\theta}{\sqrt{1-\nu}};\\
\label{dima-new-9}
\tanh 2\theta=&~-\frac{4\pi\nu}{\sqrt{1-\nu}}\frac{\tilde w}{v_c+v_n}.
\end{align}

In contrast to all states considered above, random charge tunneling between the downstream mode $\phi_1$ and the upstream modes is irrelevant in the RG sense. This can be seen from computing the scaling dimension of the charge tunneling operators $T_e=\exp(i\phi_1)\Psi_e$, where $\Psi_e$ is an electron operator that places an electron charge into the mode 
$\phi_\rho$ and excites neutral modes in the $L_n$ sector. Such random tunneling is inevitably present on the edge and establishes charge equilibration of $\phi_1$ and $\phi_\rho$. We will assume that the temperature is low and that the renormalized amplitude $A(\ell_{\rm th})$ of such random tunneling operators is small at the thermal length $\ell_{\rm th}$. The voltage 
equilibration length is the scale on which the charge exchange is significant between different channels. It is thus much longer than $\ell_{\rm th}$. We will assume below that both terminals have the same chemical potential so that the voltage is equilibrium despite a long voltage equilibration scale. 

Operators that transfer charge between modes also transfer energy. 
We thus have to consider the effect of irrelevant operators that transfer charge $me$ between upstream and downstream modes on energy equilibration. For example, one electron charge is transferred by $\hat O=\exp(i\phi_n)\psi_n$, where $\psi_n$ acts in the $L_n$ sector and has the scaling dimension \cite{Kane-disorder-dominated} $\Delta_\psi=p-\nu/[2(1-\nu)]$. The scaling dimension of $\hat O$ depends on $\theta$, Eq. (\ref{dima-new-9}), and is bounded from below  
by $p\ge 2$. The most relevant 
equilibration operator $\partial_x\phi_c\partial_x\phi_n$ transfers zero charge and has the scaling dimension $\Delta=2$ so that $\ell_{\rm eq}\sim 1/T^2$.

The filling factors $\nu=1-\frac{n}{2np-1}$, $p>1$ pose a greater challenge and do not always exhibit universal behavior. 
We set $n>1$ since $\nu=1-\frac{1}{2p-1}=1-\frac{1}{2(p-1)+1}$ were addressed above.
The action is similar to (\ref{dima-new-5}-\ref{dima-new-9}) but now $L_n$ describes $(n-1)$ downstream modes and only $\phi_n$ (\ref{dima-new-8}) runs upstream. The only upstream mode is not much faster than the rest of the modes and hence the charge transfer operators, responsible for the equilibration of the $L_n$ sector, may or may not be relevant. We will assume that they are relevant so that the Kane-Fisher mechanism \cite{Kane-disorder-dominated} makes equal the mode velocities in the $L_n$ sector. The same analysis as at $\nu=1-\frac{n}{2np+1}$ predicts then $\ell_{\rm eq}\sim 1/T^2$ at $p>2$. 

The answer changes at $p=2$, $n>1$. The most important operator $\hat O=\exp(i\phi_n)\psi_n$, where $\psi_n$ acts in the $L_n$ sector, transfers a single electron from the $\phi_1$
channel. The scaling dimension of $\psi_n$ is known \cite{Kane-disorder-dominated} and equals  $1/2-1/2n$. Hence, the scaling dimension of $\hat O$, $\Delta=2-1/n+O(\tilde w^2)$, where $O(\tilde w^2)$ is a small positive correction. We discover $\ell_{\rm eq}\sim 1/T^\alpha$ with a non-universal $\alpha\approx 2-2/n$. 

\textbf{Experimental setup} is sketched in Fig. 1. Thermal transport along the lower edge of the arm labelled $1_d=1_u$ is probed. The edge connects two floating contacts at the temperatures $T_d$ and $T_u$ and the same voltage $V$. To heat the contacts we use Joule heat \cite{heating}. Each floating contact is connected to several FQH arms. Arms $2_d$ and $2_u$
are connected to sources at the voltages $U_{d,u}$. All other arms $3_d,\dots, N_d$ and $3_u,\dots,N_u$ are grounded. The balance of the currents, entering and leaving the floating contacts, yields $G_0U_d=N_dG_0 V$ and $G_0U_u+G_0V=N_uG_0V$, where $G_0$ is the conductance of an arm. This defines the voltages $U_{d,u}$. The Joule heat dissipated in each contact is $Q_d=G_0(U_d^2-N_dV^2)/2$, $Q_u=G_0(U_u^2+V^2-N_u V^2)/2$. If needed, the dissipated heat can be changed continuously by partially closing point contact QPC$_d$. We assume that all edges, except the lower edge of arm $1_d=1_u$, are so long 
that they reach thermal equilibrium with the thermal conductance $KT=|K_u-K_d|T$,  where $K_{d,u}T$ are the combined thermal conductances of the downstream and upstream modes respectively. We chose $K_u>K_d$. Changes are minor in the opposite case.
The temperatures $T_d$ and $T_u$ can be found from Nyquist noise measurements and satisfy the energy balance equations

\begin{eqnarray}
\label{dima-new-10}
Q_d=(N_d-1)\frac{K(T_d^2-T_0^2)}{2}+\frac{(K+K_d)T_d^2-K_uT^2_u}{2}+\mathcal{J}L,\\
\label{dima-new-11}
Q_u=(N_u-1)\frac{K(T_u^2-T_0^2)}{2}-\frac{KT^2_0+K_dT_d^2-K_uT^2_u}{2}-\mathcal{J}L,
\end{eqnarray}
where $T_0$ is the ambient temperature and $\mathcal{J}$ the thermal flux (\ref{dima-old-3}) between the upstream and downstream modes on the lower edge of arm $1_d=1_u$.  Extracting $\mathcal{J}$ from the above equations, one finds the equilibration length. 

The above discussion assumed ideal contacts: normal modes leave contacts at their temperatures. To handle possible non-ideality, QPC$_L$ allows changing the length $L$ of edge $1_d=1_u$. The equilibration length can be extracted from the $L$-dependence of the data.

\textbf{Conclusions} - A promising place to look for non-Abelian orders is $\nu>2$. In particular, the states of the sixteen-fold way were proposed at $\nu=5/2$. An anti-RR order is a candidate at $\nu=12/5$. Negative-flux RR orders 
might explain the observed $\nu=12/5$ and $19/8$ ~\cite{FQH19/8_1, FQH19/8_2}. Our results at $\nu<1$ can easily be extended to higher $\nu$. The reason consists in partial decoupling of the integer and fractional channels \cite{Ken2019}. 
At sufficiently short scales only the latter participate in equilibration and their equilibration length exhibits the same temperature dependence as at
the filling factor $\nu-2$. A much longer equilibration length of the fractional and integer channels will be addressed elsewhere.

In summary, we propose a way to probe the thermal equilibration length on the edges of FQH liquids. The same universal $1/T^2$ behavior is present in most topological orders. Some orders exhibit the $1/T^4$ law. All those orders are non-Abelian and thus, the unusually fast $\sim 1/T^4$ growth of the equilibration length at low temperatures
would constitute evidence of non-Abelian statistics.

We thank A. Gromov and A. Jevicki for useful discussions. This research was supported in part by the National Science Foundation under Grant No. DMR-1902356 and Grant No. NSF PHY-1748958.

\pagebreak
\widetext

\begin{center}
\textbf{\large Supplemental Material for "Thermal equilibration on the edges of topological liquids"}
\end{center}
\setcounter{equation}{0}
\setcounter{figure}{0}
\setcounter{table}{0}
\setcounter{page}{1}
\makeatletter
\renewcommand{\theequation}{S\arabic{equation}}

\section{Equilibration length in negative-flux Read-Rezayi states} 

We need to find the most relevant operator that couples the upstream and downstream modes.
The downstream sector contains a single Bose charge mode. The upstream sector is described by the SU$(2)_k/$U$(1)$ coset chiral CFT \cite{dima-SB}. Our goal is to find in the neutral sector the most relevant Bose field that is mutually local with all primary fields.

There are $k(k+1)/2$ primary fields $\Phi^j_m$ in the neutral sector, where $j=0,1/2,\dots,k/2$, 
$(j-m)\in\mathbb{Z}$, and two identifications are made: $(j,m)\equiv(j,m+k)$ and $(j,m)\equiv (\frac{k}{2}-j,m+\frac{k}{2})$. The identifications allow choosing $j>0$ and $-j<m\le j$ for each primary field. With such choice, the scaling dimension of $\Phi^j_m$ expresses as

\begin{equation}
\label{dima-new-3}
\Delta^j_m=\frac{j(j+1)}{k+2}-\frac{m^2}{k}.
\end{equation}
A different choice of the indices for the same field would require adding an integer to the right hand side of the above equation. The fusion rules are

\begin{equation}
\label{dima-new-4}
\Phi^j_m\times\Phi^{j'}_{m'}=\sum_{j''=|j-j'|}^{{\rm min}(j+j',k-j-j')}\Phi^{j''}_{m+m'}.
\end{equation}
The electron operator is $\Phi^{k/2}_{1-k/2}\exp(i\phi_c/\nu)$. The simplest quasiparticle operator $\sigma=\Phi^{1/2}_{1/2}\exp(i[k-1]\phi_c/k)$.

To construct operators $\hat O$, responsible for equilibration, we must identify the most relevant bosonic operator in the neutral sector. We start with the primary fields. One Bose field is immediately apparent: it is $\Phi^{k/2}_{k/2}$ with $\Delta^{k/2}_{k/2}=0$. Of course, the scaling dimension shows that the field is a trivial constant, useless for our purposes. We now show that no other primary Bose field $\Phi^j_m$ exists. Since we look for a field from the vacuum superselection sector, it should have only one fusion channel with $\sigma$.  This implies  $j=k/2$. The fusion outcome is $f=\Phi^{(k-1)/2}_{m+1/2}\exp(i[k-1]\phi_c/k)$. Second, fusing with the Bose field $\Phi^{k/2}_m$ should not change the superselection sector of $\sigma$.
In particular, it does not change the topological spin $\exp(\pi i \nu [1-1/k]^2 - 2\pi i \Delta^{1/2}_{1/2})$.  Hence, the difference of the scaling dimensions $\Delta_f-\Delta_\sigma=(\Delta^{k/2}_m-\frac{1}{2}-\frac{m}{k})~{\rm mod}~\mathbb{Z}$ must be an integer. Observing that the scaling dimension $\Delta^{k/2}_m$ of a Bose field is also an integer, we discover a single possibility
 $m=j=k/2$ that was already addressed.

We thus have to examine descendant fields, which are obtained \cite{cft} from the primary fields by the action of conformal generators $\hat L_{-n}$. 
We are only interested in the descendants of the trivial constant $\Phi^{k/2}_{k/2}$ since $\hat L_{-n}$ cannot change the superselection sector. The only candidate for the scaling dimension of 1 is $\hat L_{-1}\Phi^{k/2}_{k/2}=\partial~{\rm const}=0$. Thus, the maximal possible scaling dimension of a non-trivial Bose-operator in the neutral sector is 2. Such operator exists: it is the energy-momentum tensor $\hat T$. Hence, the most relevant operator in Eq.~\eqref{dima-2} is $\hat O=\partial_x\phi_c \hat T$ with the scaling dimension $\Delta=3$. We find $\ell_{\rm eq}\sim1/T^4$. 


\begin{thebibliography}{99}


\bibitem{Wen_book}
X.-G. Wen, \textit{Quantum Field Theory of Many-Body Systems: From the Origin of Sound to an Origin of Light and Electrons} (Oxford University Press, 2004).

\bibitem{kitaev2006}
A. Kitaev,
\href{https://www.sciencedirect.com/science/article/pii/S0003491605002381?via\%3Dihub}
 {Annals of Physics \textbf{321}, 2 (2006)}.
 
\bibitem{16-fold}
K. K. W. Ma and D. E. Feldman,
\href{https://journals.aps.org/prb/abstract/10.1103/PhysRevB.100.035302}
{Phys. Rev. B \textbf{100}, 035302 (2019)}.

\bibitem{MFreview}
M. Heiblum and D. E. Feldman
\href{https://arxiv.org/abs/1910.07046}
{arXiv:1910.07046}.

\bibitem{Kane_thermal}
C. L. Kane and M. P. A. Fisher,
\href{https://journals.aps.org/prb/abstract/10.1103/PhysRevB.55.15832}
{Phys. Rev. B \textbf{55}, 15832 (1997)}.

\bibitem{Read-Green}
N. Read and D. Green,
\href{https://journals.aps.org/prb/abstract/10.1103/PhysRevB.61.10267}
{Phys. Rev. B \textbf{61}, 10267 (2000)}.

\bibitem{Cappeli}
A. Cappelli, M. Huerta, and G. R. Zemba,
\href{https://www.sciencedirect.com/science/article/pii/S0550321302003401?via%3Dihub}
{Nucl. Phys. B \textbf{636}, 568 (2002)}.

\bibitem{Banerjee2017}
M. Banerjee, M. Heiblum, A. Rosenblatt, Y. Oreg, D. E. Feldman, A. Stern, and V. Umansky,
\href{https://www.nature.com/articles/nature22052}
{Nature \textbf{545}, 75 (2017)}.

\bibitem{Banerjee2018}
M. Banerjee, M. Heiblum, V. Umansky, D. E. Feldman, Y. Oreg, and A. Stern, 
\href{https://www.nature.com/articles/s41586-018-0184-1}
{Nature \textbf{559}, 205 (2018)}.

\bibitem{RuCl3}
Y. Kasahara, T. Ohnishi, Y. Mizukami, O. Tanaka, S. Ma, K. Sugii, N. Kurita, H. Tanaka, J. Nasu, Y. Motome, T. Shibauchi, and Y. Matsuda, 
\href{https://www.nature.com/articles/s41586-018-0274-0} 
{Nature (London) {\bf 559}, 227 (2018)}.


\bibitem{so-2018}
A. Aharon-Steinberg, Y. Oreg, and A. Stern,
\href{https://link.aps.org/doi/10.1103/PhysRevB.99.041302}
{Phys. Rev. B \textbf{99}, 041302(R) (2019)}.

\bibitem{Ken2019}
K. K. W. Ma and D. E. Feldman,
\href{https://journals.aps.org/prb/abstract/10.1103/PhysRevB.99.085309}
{Phys. Rev. B \textbf{99}, 085309 (2019)}.


\bibitem{Lafont2019}
F. Lafont, A. Rosenblatt, M. Heiblum, and V. Umansky,
\href{http://science.sciencemag.org/content/363/6422/54.editor-summary}
{Science \textbf{363}, 54 (2019)}.

\bibitem{Son2015}
D. T. Son,
\href{https://journals.aps.org/prx/abstract/10.1103/PhysRevX.5.031027}
{Phys. Rev. X \textbf{5}, 031027 (2015)}.

\bibitem{Zucker2016}
P. T. Zucker and D. E. Feldman,
\href{https://journals.aps.org/prl/abstract/10.1103/PhysRevLett.117.096802}
{Phys. Rev. Lett. \textbf{117}, 096802 (2016)}.

\bibitem{APf_Lee2007}
S.-S. Lee, S. Ryu, C. Nayak, and M. P. A. Fisher,
\href{https://journals.aps.org/prl/abstract/10.1103/PhysRevLett.99.236807}
{Phys. Rev. Lett. \textbf{99}, 236807 (2007)}.

\bibitem{FCV}
L. Fidkowski, X. Chen, and A. Vishwanath, 
\href{https://journals.aps.org/prx/abstract/10.1103/PhysRevX.3.041016}
{Phys. Rev. X \textbf{3}, 041016 (2013)}.

\bibitem{BNQ}
P. Bonderson, C. Nayak, and X.-L. Qi, 
\href{https://iopscience.iop.org/article/10.1088/1742-5468/2013/09/P09016}
{J. Stat. Mech. \textbf{2013}, P09016}.

\bibitem{Jolicoeur2007}
Th. Jolicoeur,
\href{https://journals.aps.org/prl/abstract/10.1103/PhysRevLett.99.036805}
{Phys. Rev. Lett. \textbf{99}, 036805 (2007)}.

\bibitem{RR}
N. Read and E. Rezayi,
\href{https://journals.aps.org/prb/abstract/10.1103/PhysRevB.59.8084}
{Phys. Rev. B \textbf{59}, 8084 (1999)}.


\bibitem{Kane-disorder-dominated}
C. L. Kane and M. P. A. Fisher,
\href{https://journals.aps.org/prb/abstract/10.1103/PhysRevB.51.13449}
{Phys. Rev. B \textbf{51}, 13449 (1995)}.

\bibitem{dima-footnote-03-19}
The exponent in Eq. (\ref{eq:l-eff-RG}) ends up being $3-2\Delta$.

\bibitem{Dima_comment2018}
D. E. Feldman,
\href{https://journals.aps.org/prb/abstract/10.1103/PhysRevB.98.167401}
{Phys. Rev. B \textbf{98}, 167401 (2018)}.

\bibitem{Jain_book}
J. K. Jain, \textit{Composite Fermions} (Cambridge University Press, Cambridge, 2007).


\bibitem{foot2}
It is essential to switch to the weakly interacting Kane-Fisher variables in the discussion of thermal transport since different temperatures cannot be assigned to strongly interacting modes.

\bibitem{APf_Levin2007}
M. Levin, B. I. Halperin, and B. Rosenow,
\href{https://journals.aps.org/prl/abstract/10.1103/PhysRevLett.99.236806}
 {Phys. Rev. Lett. \textbf{99}, 236806 (2007)}.


\bibitem{guang2013}
G. Yang and D. E. Feldman,
\href{https://journals.aps.org/prb/abstract/10.1103/PhysRevB.88.085317}
{Phys. Rev. B \textbf{88}, 085317 (2013)}.

 \bibitem{guang2014}
 G. Yang and D. E. Feldman,
 \href{https://journals.aps.org/prb/abstract/10.1103/PhysRevB.90.161306}
{Phys. Rev. B \textbf{90}, 161306(R) (2014)}.



\bibitem{Bishara_PH_RR}
W. Bishara, G. A. Fiete, and C. Nayak,
\href{https://journals.aps.org/prb/abstract/10.1103/PhysRevB.77.241306}
{Phys. Rev. B \textbf{77}, 241306(R) (2008)}.

\bibitem{cft}
P. Di Francesco, P. Mathieu, and D. Senechal, \textit{Conformal Field Theory} (Springer-Verlag, New York, 1997).

\bibitem{foot3}
It is worthwhile to examine the role of non-random interactions in the PH-Pfaffian state.  In all states we considered before the PH-Pfaffian order, the most important equilibration operator was random, $\int dx \xi(x)\hat O(x)$. A non-random operator $\xi\int dx \hat O(x)$ with the same $\hat O(x)$ was prohibited by symmetry. No such prohibition exists for the operator
$i \xi\int dx \partial\phi_c\psi\partial_x\psi$ in the PH-Pfaffian case. Naive scaling analysis then suggests the equilibration length $\sim 1/T^3$. The naive calculation fails since simultaneous energy and momentum conservation is impossible in a translationally invariant 1D system with only two different mode velocities. Curvature of the spectrum may lift this constraint
but introduces additional powers of $1/T$ in the estimate for $\ell_{\rm eq}$. A subtlety involves approximate momentum conservation in a finite system: the accuracy is $\sim \hbar/L$. The validity of our results then requires $\frac{\rm const}{T^2}\ll L$ in the PH-Pfaffian case.

\bibitem{dima-SB}
The discussion from
[J. K. Slingerland and F. A. Bais,
\href{https://www.sciencedirect.com/science/article/pii/S055032130100308X?via\%3Dihub} 
 {Nucl. Phys. B \textbf{612}, 229 (2001) }] is particularly convenient for our purposes.
 
\bibitem{SM-dima} See Supplemental Material for the discussion of thermal equilibration length in negative-flux Read-Rezayi states.

\bibitem{heating}
S. Jezouin, F. D. Parmentier, A. Anthore, U. Gennser, A. Cavanna, Y. Jin, and F. Pierre, 
\href{https://science.sciencemag.org/content/342/6158/601}
{Science \textbf{342}, 601 (2013)}.


\bibitem{FQH19/8_1}
J. S. Xia, W. Pan, C. L. Vicente, E. D. Adams, N. S. Sullivan, H. L. Stormer, D. C. Tsui, L. N. Pfeiffer, K. W. Baldwin, and K. W. West,
\href{https://journals.aps.org/prl/abstract/10.1103/PhysRevLett.93.176809}
{Phys. Rev. Lett. \textbf{93}, 176809 (2004)}.

\bibitem{FQH19/8_2}
W. Pan, J. S. Xia, H. L. Stormer, D. C. Tsui, C. Vicente, E. D. Adams, N. S. Sullivan, L. N. Pfeiffer, K. W. Baldwin, and K. W. West,
\href{https://journals.aps.org/prb/abstract/10.1103/PhysRevB.77.075307}
{Phys. Rev. B \textbf{77}, 075307 (2008)}.


\end{thebibliography}
\end{document}